# The Effects of Electronic and Photonic Coupling on the Performance of a Photothermionic-Photovoltaic Hybrid Solar Device

Ehsanur Rahman[1,2]* and Alireza Nojeh[1,2]

[1] Department of Electrical and Computer Engineering, University of British Columbia, Vancouver, BC, V6T 1Z4, Canada

[2] Quantum Matter Institute, University of British Columbia, Vancouver, BC, V6T 1Z4, Canada

*Email: ehsanece@ece.ubc.ca

## Abstract

This work presents a detailed analysis of the photothermionic-photovoltaic hybrid solar device. The electrons in this hybrid device gain energy from both the solar photons and thermophotons generated within the device, and hence the device has the potential to offer a voltage boost compared to individual photothermionic or photovoltaic devices. We show that the gap size between the photothermionic emitter and the photovoltaic collector crucially affects the device performance due to the strong dependence of the electronic and photonic coupling strengths on this gap size. We also investigate how the current matching constraint between the thermionic and photovoltaic stages can affect the hybrid solar device performance by studying different hybrid device configurations. Moreover, the hybrid devices are compared with the single photothermionic solar device with a metallic collector. Interestingly, we observe that the addition of a photovoltaic stage meant to enable the hybrid device to capture the entire terrestrial solar spectrum does not necessarily lead to higher overall conversion efficiency.

**Keywords**

Photothermionic, Photovoltaic, Hybrid solar device, Space charge, Near-field effect.

## 1. Introduction

**E**fficient harvesting of solar energy is crucial due to its inexhaustible and clean nature, as well as widespread availability. Established solar harvesting technologies exploit either the thermal or optical nature of solar energy. For example, the photovoltaic (PV) effect, which relies on optical excitation, is a prominent mechanism of solar electricity generation. However, the thermal impact is detrimental to photovoltaic devices' operation [1]. Also, complete utilization of the solar spectrum is difficult [2,3]. Multijunction photovoltaic (MPV) devices suffer from lattice and current mismatch among the epitaxial layers with different bandgaps [4]. Concentrated photovoltaic (CPV) devices involve demanding heat management for efficient operation [5,6]. Concentrated solar thermal power (CSP) plants require large infrastructure (*e.g.,* heliostat, parabolic trough reflectors, linear Fresnel concentrators, etc.) for collecting and delivering the sunlight to a central solar receiver tower and absorber tube containing the heat transfer fluids [7,8]. In CSP plants, there are inevitable thermal losses during the transfer of the absorbed solar thermal power to the conversion plant. The heat transfer fluids are also limited in terms of their maximum operating temperature, thereby limiting the efficiency of the conversion process, and they might also cause corrosion of the metal housing [9].

A relatively new solar conversion approach is based on photon enhanced thermionic emission [10], which relies on thermal excitation for electron emission. (For the sake of brevity, we define a device based on this mechanism as a photothermionic (PT) device.) Similarly to the PV effect, this approach exploits the quantum nature of solar photons by generating electron-hole pairs. On the other hand, like CSP plants, it also utilizes any excess energy (which is lost by the thermalization process in the PV mechanism) of the absorbed above-bandgap solar photons [11]. However, PT devices have their own limitations. For example, the space charge effect is a key challenge [12,13], which can be mitigated by using a microscale gap between the device electrodes. At such a small gap scale, thermal radiation loss can be crucial due to the coupling of evanescent photons (also known as the near-field effect) and limits the performance of the device [14,15]. Moreover, similar to photovoltaics, sub-bandgap photons are not utilized in the conversion process [16]. Therefore, the conversion performance of a PT solar device may in principle be improved if both the unabsorbed solar photons and the thermophotons generated by the PT emitter are captured by a second, PV stage.

Motivated by the above prospects, in this work, we investigate a hybrid solar device that combines the PT and PV technologies into a single device with different possible electronic coupling configurations. The basic concept of combining the thermionic and photovoltaic mechanisms is not new and has been previously proposed for near-field thermophotovoltaic devices [17,18]: the resulting hybrid device is also known as a thermionic enhanced thermophotovoltaic device (TI-PV). In such a device, the thermal radiator is replaced with metal. The motivations behind using a metallic thermal radiator are two-fold. Firstly, at the operating temperature of the device, in addition to photons, the metal could also supply energetic electrons via thermionic emission. These high-energy electrons can generate a higher output voltage compared to that of the individual thermophotovoltaic (TPV) device. Secondly, the thermionically emitted electrons recombine with the photo-generated holes in the TPV device thereby creating a direct electronic coupling between the radiator and the TPV device. As a result, the TPV sub-device in this hybrid device can operate without a front electrode and does not suffer from the losses (*e.g.,* optical shading and voltage drop in the series resistance) associated with such a front electrode. Moreover, in a near-field TPV device, the strength of the radiative coupling between the thermal radiator and the PV device crucially depends on the gap size between them [19–21], and the minimum achievable gap size may be limited by the thickness of the front electrode [17]. The TI-PV architecture also eliminates this constraint. Due to these reasons, the TI-PV architecture has the potential to further improve the performance of the TPV device. However, the TI-PV device requires current matching between the thermionic emitter and the photovoltaic sub-devices [17,22]; the current from either stage could be a limiting factor in exploiting the full potential of both the thermionic and photovoltaic stages. For example, in a recent experimental implementation of such TI-PV devices [23], although the TPV sub-device was shown to gain an additional voltage from the thermionic stage, it was biased near the open-circuit regime due to the limited current available from the thermionic emitter. This issue can be solved by independently biasing the thermionic and photovoltaic sub-devices [24]. However, in the latter configuration, the PV sub-device may suffer from the resistive and shading losses associated with the additional electrode deposited on its top surface. These challenges are also relevant to the photothermionic-photovoltaic (PT-PV) solar devices presented in this work. Therefore, although such a hybrid solar conversion technology is conceptually interesting, optimizing its operation is not trivial and a systematic understanding of the various physics that might be limiting its performance is needed.

In this work, we analyze the operation of different hybrid PT-PV solar devices and compare their performance with that of an individual photothermionic device. We show how the electronic and photonic coupling between the PT emitter and the PV device affect the hybrid device operation and quantify the performance gain from such a hybrid device compared to a PT device. Therefore, this work sheds light on the intricate physics involved in the operation of the photothermionic-photovoltaic solar device and our findings can guide the design of such a hybrid device for experimental implementation.

**2. Device Operation**

We first briefly describe the operation of a PT-PV hybrid solar device and its different configurations with the help of simple device schematics and energy band diagrams as shown in fig. 1(a)-(d). The hybrid device consists of a semiconductor-based photothermionic emitter (or cathode) which absorbs the incident solar photons with energy higher than its bandgap. The photogenerated electrons thermalize and may undergo various recombination processes during their transport towards the emitting surface. Under appropriate conditions, these excess electrons may also cause a quasi-Fermi level splitting within the emitter band structure and help reduce the effective work function of the thermionic emitter; this is known as the photon enhancement effect [25]. The PV sub-device is made of a low-bandgap semiconductor material so that it can absorb both the thermophotons generated by the emitter as well as the solar photons which are not absorbed by the emitter. For example, if the emitter has a bandgap of 1eV, 15% of the incident solar photons will cross the silicon emitter. The photons which are absorbed by the PV sub-device further upshift the electron Fermi level and generate an additional useful voltage.

Depending on the device structure, the PV sub-device is either forced to be biased at the same current density as the PT emitter (fig. 1a), resulting in a two-terminal configuration, or allowed to be biased independently (fig. 1b), resulting in a three-terminal configuration. In the former configuration (*i.e.,* the coupled or two-terminal device), the thermionic electrons leaving the photothermionic emitter will be absorbed by the PV sub-device's top layer where they will recombine with the photogenerated holes [17], thereby maintaining the same current through the PV sub-device. In the latter configuration, a metallic grid is deposited on the top layer of the PV sub-device so that the current in the thermionic and photovoltaic sub-devices can be biased separately. In this configuration, the thermionic and PV stages are electronically decoupled. In both configurations, the PV top layer is coated with a very thin low work function material for efficient collection of the thermionic electrons. Due to its very small thickness, this coating layer can be optically transparent [18,23,26]. However, in the second configuration (fig. 1b), the grid electrode will also cause some optical shading loss. Also, the thickness of the grid would limit the minimum achievable gap size between the PT emitter and the PV sub-device.

To overcome the shading loss, instead of the grid, a thin transparent conductive layer (*e.g.*, transparent conductive oxide films or graphene) may be used [27]. However, previous experimental implementations of such a transparent contact layer using graphene have shown very high sheet resistance values (of the order of kΩ/sq), given the tradeoff between the graphene stack's transparency and sheet resistance [28–31]. On the other hand, indium-tin-oxide front contacts in silicon and other thin-film solar devices have been reported to have a sheet resistance of the order of a few tens of Ω/sq [28,32], but even such low values of sheet resistance might result in a negligible PV output voltage under the device operating conditions considered in this study. Therefore, to evaluate the performance of the decoupled device, here we have considered two different levels of realism.

In one case, we model the effects of the grid electrode more realistically by assuming a finite grid thickness and taking into account the associated optical shading loss caused by the grid. In this case, for a given series resistance value, a thicker grid would require lower areal coverage and thus result in lower optical shading loss, and vice versa [33]. However, a very thick grid would significantly limit the minimum achievable distance between the PT emitter and the PV top surface, possibly resulting in a severe space charge loss in the vacuum gap [34]. For reasonable thermionic conversion efficiency, a vacuum gap of a few micrometers is required [14,15,35]; therefore, here we use a grid thickness of 1 μm. Aiming for a series resistance value of 10 mΩ, we expect a grid shading loss of around 20% [17]. In another case, simply to estimate the upper limit of performance of the decoupled PT-PV hybrid solar device, we assume that the PV sub-device performance is not limited by the optical shading and thickness of the grid (*i.e.*, we assume a grid-less contact with a series resistance value of 10 mΩ).

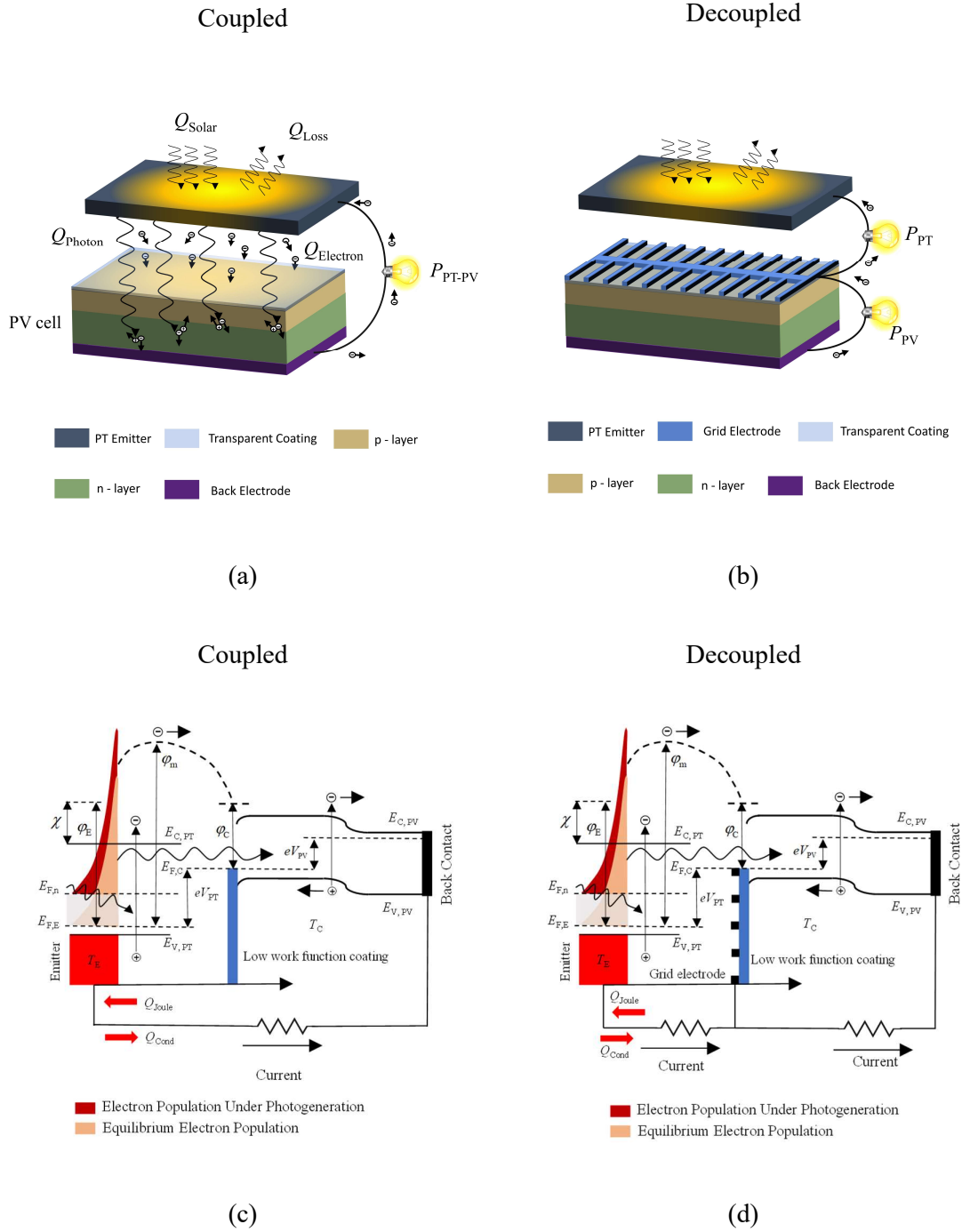

**Fig. 1.** (a)-(b) Simple device schematics and (c)-(d) corresponding band diagrams of various PT-PV hybrid solar devices. In the band diagrams, $E_{C,\,PT}$ and $E_{C,\,PV}$ are the conduction band minima of the PT and PV sub-devices, respectively. Similarly, $E_{V,\,PT}$ and $E_{V,\,PV}$ are the valance band maxima of the PT and PV sub-devices, respectively.

$\chi$ is the electron affinity at the PT emitting surface. $\varphi_E$ and $\varphi_C$ are the work functions of the PT emitting surface and low work function coating layer, respectively and $\varphi_m$ is the vacuum barrier for thermionic emission. $E_{F,E}$ and $E_{F,n}$ are the equilibrium and steady-state non-equilibrium Fermi levels in the PT emitter, respectively, and $E_{F,C}$ is the Fermi level of the collector. $V_{PT}$ and $V_{PV}$ are the voltages of the PT and PV sub-devices respectively, and $e$ is the electron charge. $T_E$ and $T_C$ are the temperature of the PT emitter and PV device, respectively. Panels (a) and (c) correspond to the coupled device and panels (b) and (d) correspond to the decoupled device. In the coupled device, the photothermionic emitter and the photovoltaic sub-device are connected in series whereas, in the decoupled device, the photothermionic and photovoltaic sub-devices are biased independently.

## 3. Methodology

To analyze the hybrid PT-PV solar device, we have considered various relevant physics in the thermionic and photovoltaic sub-devices. The theories we used to capture these physics are well established and have been previously used to analyze individual photothermionic and photovoltaic solar devices. Here we provide a brief overview of these theories and refer to the literature where the computational details can be found.

### 3.1. Photothermionic Emitter Model

For the analysis of the photothermionic emitter, we need the spatial distribution of the carriers inside the emitter, which can be obtained by solving the particle continuity equation:

$$\vec{\nabla}.\vec{J_e} = -e(G-R), \quad (1)$$
$$\vec{\nabla}.\vec{J_h} = e(G-R), \quad (2)$$

where
$$\vec{J_e} = e\mu_e n \vec{\nabla}\varphi_e, \quad (3)$$
$$\vec{J_h} = e\mu_h p \vec{\nabla}\varphi_h. \quad (4)$$

In the above, $\vec{J_e}$ and $J_h$ are the electron and hole current densities, respectively; $\mu_e$ and $\mu_h$ are the electron and hole mobilities, respectively; $n$ and $p$ are the steady-state electron and hole concentrations, respectively; $\varphi_e$, $\varphi_h$ are the electron and hole quasi-Fermi potentials, respectively; $G$ and $R$ are the local photogeneration and recombination rates of electron-hole pairs (EHPs), respectively. To obtain $G$, we considered the spatial dependence of optical absorption in the PT emitter [36]. To obtain $R$, we implemented the radiative, surface, Auger and Shockley-Reed-Hall recombination processes using theories valid under both low and high injection levels [37]. The detailed implementation of the emitter model can be found in [11].

### 3.2. Electronic Coupling Model

The strength of electronic coupling between the PT emitter and collector was calculated considering the space charge opposing the thermionically emitted electrons. As they traverse the vacuum gap towards the collector, these electrons may have to overcome an additional potential barrier due to their mutual coulombic repulsion, which is known as the space charge effect [34]. To incorporate this effect, we have calculated the velocity distribution of the electrons using the Vlasov equation [38] as

$$f(x,v_e) = 2n(x_m)\sqrt{m_e^3/(8\pi^3 k_B^3 T_E^3)} \exp\left(\frac{\varphi_m - \varphi(x) - \frac{1}{2}m_e v_e^2}{k_B T_E}\right) \Theta(v_{ex} \mp v_{ex,min}), \quad (5)$$

where $\varphi_m$ and $\varphi_x$ are the maximum and local potential barriers in the interelectrode space, respectively, $m_e$ is the electron mass, $k_B$ is the Boltzmann constant, $T_E$ is the emitter temperature, $v_e$ and $v_{ex}$ are the electron velocity and its component perpendicular to the emitting surface, respectively, $n(x_m)$ is the electron density at the position of maximum motive, and $\Theta$ is the Heaviside step function. The upper and lower signs apply for $x > x_m$ and $x \leq x_m$, respectively. The $v_{ex,min}$ is defined as $v_{ex,min} = -\sqrt{2(\varphi_m - \varphi(x))/m_e}$ where $\varphi_m$ is the maximum motive in the interelectrode space, $\varphi(x)$ is the local motive and $m_e$ is the electron mass. We note that the image charge effect may in principle affect the potential barrier to electron transport; however, this effect is expected to be insignificant [39].

By integrating the electron distribution over the entire velocity space, we obtain the spatial distribution of charge carriers inside the vacuum gap as [38]

$$n(x) = \int_{-\infty}^{+\infty} dv_z \int_{-\infty}^{+\infty} dv_y \int_{-\infty}^{+\infty} dv_x f(x, v_e) = n(x_m) \exp(\gamma)[1 \mp \text{erf}(\sqrt{\gamma})], \quad (6)$$

where $\gamma = (\varphi_m - \varphi(x))/k_B T_E$ is the dimensionless potential barrier and $\text{erf}(z) = \frac{2}{\sqrt{\pi}} \int_0^z \exp(-t^2) dt$ is the error function. The upper and lower signs apply for $x > x_m$ and $x \leq x_m$, respectively. The resulting space charge barrier is calculated by solving the Poisson equation, which can be written as

$$2 \frac{d^2 \gamma}{d\xi^2} = \exp(\gamma)[1 \pm \text{erf}(\sqrt{\gamma})], \quad (7)$$

where $\xi$ is the dimensionless position variable given by $\xi = (x - x_m)/x_0$, where $x_0$ is the normalization length. The upper and lower signs apply for $x < x_m$ and $x \geq x_m$, respectively. The details regarding the implementation of the above steps can be found in [40].

By solving the above, we obtain the potential profile in the interelectrode space as well as the maximum value of the potential barrier. The space charge limited current from the emitter can be written as

$$J_E = A_R T_E^2 \exp[-(\varphi_m - E_{F,n} + E_{F,eq})/k_B T_E], \quad (8)$$

$$\text{where } E_{F,n} - E_{F,eq} = k_B T_E \ln(n/n_{eq}). \quad (9)$$

In the above equations, $E_{F,n}$ and $E_{F,eq}$ are the steady-state and equilibrium electron Fermi levels, respectively, $A_R$ is the Richardson constant and $n$, $n_{eq}$ are the steady-state and equilibrium electron density in the PT emitter, respectively. We also note that we have assumed no electron reflection from the collector. This is justified based on our previous results [14] showing that a change in the reflection coefficient does not lead to a qualitative change in the outcome.

### 3.3. Photonic Coupling Model

The photonic coupling between the photothermionic emitter and the photovoltaic sub-device consists of the near-field evanescent photons as well as the far-field propagating photons, and the possible interference due to the wave nature of these photons in the vacuum gap. The physics of the photonic coupling (considering the complete solar device structure) is captured by fluctuational electrodynamics [41]. In brief, the origin of thermal radiation from a medium at a finite temperature can be traced to the random thermal fluctuations of charges or dipoles contained within that medium [42]. The energy flux in the resulting fluctuating electromagnetic fields is calculated from the time-averaged Poynting vector component perpendicular to the emitting surface. The time-averaged Poynting vector is obtained by solving the stochastic Maxwell's equations and can be written as [41]

$$\langle \vec{S}(\vec{r},\omega) \rangle = 2\,\mathrm{Re}\{\langle \vec{E}(\vec{r},\omega) \times \vec{H}^*(\vec{r},\omega) \rangle\}, \quad (10)$$

where

$$\vec{E}(\vec{r},\omega) = i\omega\mu_0 \int_V \overline{\overline{G}}^E(\vec{r},\vec{r}',\omega) \cdot \vec{j}(\vec{r}',\omega) dV, \quad (11)$$

$$\vec{H}(\vec{r},\omega) = \int_V \overline{\overline{G}}^H(\vec{r},\vec{r}',\omega) \cdot \vec{j}(\vec{r}',\omega) dV. \quad (12)$$

In the above equations, $\vec{E}(\vec{r},\omega)$ and $\vec{H}(\vec{r},\omega)$ are the electric and magnetic fields at a location $\vec{r}$ due to a fluctuating current source $\vec{j}(\vec{r}',\omega)$ located at $\vec{r}'$ within the radiating medium of volume $V$. The terms $\overline{\overline{G}}^E(\vec{r}',\vec{r},\omega)$ and $\overline{\overline{G}}^H(\vec{r}',\vec{r},\omega)$ are, respectively, the electric and magnetic Dyadic Green's functions relating the field at a location $\vec{r}$ due to a current source at $\vec{r}'$. The detailed implementation of this model for layered media can be found in [43]. The factor of 2 is included in the Poynting vector calculation since only positive frequencies are considered in the Fourier decomposition of the time-dependent fields into frequency-dependent quantities [41].

### 3.4. Thermal Balance Model

The PT emitter temperature is calculated by solving for thermal balance, which considers the various processes by which the absorbed solar energy is taken away from the emitter by electrons, photons, and phonons, *i.e.*,

$$Q_{In} = Q_{Electron} + Q_{Photon} + Q_{Phonon}. \quad (13)$$

The electronic energy exchange involves the thermionic emission process and can be written as [44]

$$Q_{Electron} = \frac{[(J_E - J_C)\varphi_m + 2k_B(T_E J_E - T_C J_C)]}{e}. \quad (14)$$

The photonic energy exchange involves the thermal radiation processes. The emitter thermally radiates energy to the surroundings as well as to the PV sub-device and also receives the energy which is radiated back by them. These radiative exchanges can be enhanced due to non-equilibrium radiative recombination processes. Thermal radiative exchange with the ambient is calculated using the Stefan-Boltzmann law considering the spectral emissivity of the PT emitter material [11]. The radiative exchange between the PT emitter and the PV sub-device strongly depends on the vacuum gap width, and we calculate it using fluctuational electrodynamics as explained above. In summary,

$$Q_{\text{Photon}} = Q_{\text{Surrounding}} + Q_{\text{Interelectrode}} \,. \qquad (15)$$

In addition, in the electric lead, there are thermal conduction and Joule heating losses via phonons. This can be written as [34]

$$Q_{\text{Phonon}} = L(T_E^2 - T_C^2)/(2R_{\text{Lead}}) - (SJ)^2 R_{\text{Lead}}/2 \,, \qquad (16)$$

where $L$ is the Lorenz number for the lead material, $R_{\text{Lead}}$ is the lead resistance and $S$ is the cross-sectional area of the device.

The lead resistance needs to be optimized considering the trade-off between the above two losses, resulting in

$$\eta_{\text{Lead optimized}} = \frac{SJ_{\text{net}}(V_o - SJ_{\text{net}} R_{\text{Lead}})_{\max}}{SQ_{\text{In}}}. \qquad (17)$$

In the above, $V_o = V_{\text{PT}} + V_{\text{PV}}$ for the electronically coupled device, and $V_o = V_{\text{PT}}$ for the decoupled and single PT devices, respectively.

### 3.5. Particle Balance Model

In the photothermionic emitter, the thermionic current depends on the steady-state electron concentration at the emitting surface, emitter temperature and work function. This electron concentration is obtained by solving for particle balance at the emitting surface, but is also a strong function of the emitter temperature [45], which is obtained by solving for thermal balance. Therefore, the particle and thermal balance are coupled and need to be solved in a self-consistent iterative process, for which the algorithm is shown in fig. 2.

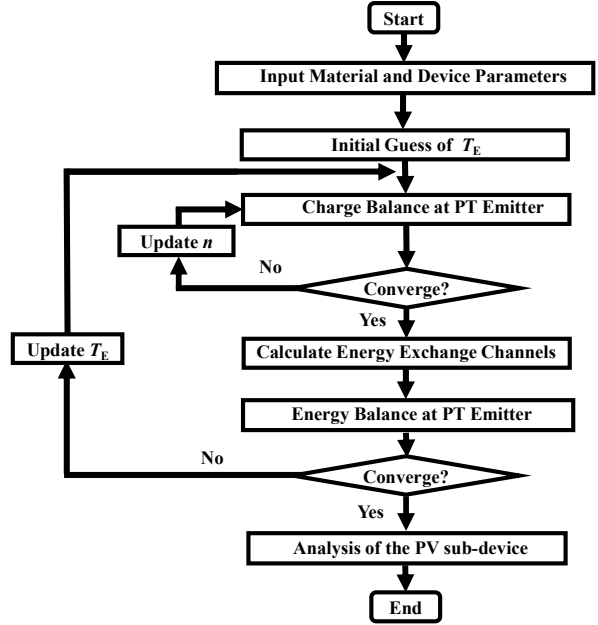

Fig. 2. A flowchart of the self-consistent numerical iterative algorithm implemented for the hybrid solar device analysis.

## 3.6. Photovoltaic Device Model

Similar to the PT emitter, the minority carrier distributions in the PV sub-device need to be obtained by solving the particle continuity equation in the p and n regions. To this end, the photon absorption profile in various regions of the PV sub-device (*i.e.,* p region, n region and the depletion layer formed at the junction between them) was calculated using fluctuational electrodynamics as described before. From this photon absorption profile, we calculate the photogeneration rate of electron-hole pairs, which is then used to calculate the spatial distribution of minority carriers in various regions of the PV sub-device by solving the particle continuity equation (as described for the PT emitter) with appropriate boundary conditions. Once the spatial distribution of the minority carriers is obtained, the photocurrent from various regions can be calculated. The dark current can be calculated similarly but assuming no photoexcitation. The computational details of the above steps can be found in [46,47]. For lowering the work function of the PV front surface, we considered a thin low work function coating with a work function of 1 eV. (This value of work function is reasonable; for example, phosphorus-doped diamond films have been shown to achieve a work function of 0.9 eV [48].) The PV sub-device temperature was kept at 300 K. Note that there could be a coupling between the radiation transfer and charge transport within the PV sub-device; however, this effect is negligible for gap sizes over 100 nm [49].

### 3.7. Material Properties

For the PT emitter material in all devices considered, we have used silicon (Si) (which is also widely used in single-junction photovoltaics) due to its high melting point as well as its bandgap, which is suitable for absorbing the solar spectrum. However, Si is an indirect bandgap material and has a weak absorption coefficient, necessitating a large thickness for efficient absorption of the solar spectrum—we use a thickness of 20 μm. (It has been observed that the conversion performance of Si photothermionic devices saturates beyond an emitter thickness of around 10 μm [11]. This is because a 10 μm thickness is nearly sufficient for the effective absorption of the portion of the solar spectrum absorbable by the Si-based emitter [12].). We considered a p-type doping level of $10^{18}$ cm$^{-3}$ with the boron dopant energy level, an electron affinity of 1 eV (obtainable through the appropriate surface coating[50]) and the theoretical value of 120 Acm$^{-2}$ K$^{-2}$ for the Richardson constant. We used the AM 1.5D incident solar spectrum [51] with the concentration ratio given later. For the analysis of the single PT solar device, we considered a collector with the dielectric properties of tungsten and a work function of 1 eV.

For the PV sub-device, we considered indium arsenide (InAs) due to its low bandgap, which is suitable for the absorption of both thermophotons (generated by the PT emitter) and the low energy solar photons (which are not absorbed by the PT emitter). For the hybrid solar devices studied in this work, the cut off frequency of the absorbable solar spectrum is determined by the bandgap of the photovoltaic sub-device (neglecting intra-bandgap absorptions). The bandgap (0.354 eV) of InAs is sufficient to absorb more than 99.5% of the AM 1.5 solar spectrum. Therefore, a lower bandgap material would not yield any noticeable advantage as far as the absorption of the solar spectrum is concerned. Moreover, a lower bandgap would lead to a reduced output voltage from the photovoltaic sub-device. As well, if we consider the room temperature operation of the PV sub-device, a lower bandgap material might not even operate in the photovoltaic mode due to the excessive thermal generation of electron-hole pairs. For example, Indium Antimonide based thermophotovoltaic devices require cryogenic cooling [52] to overcome this thermal limitation, and such cryogenic cooling would require additional power.

For the p layer, we considered a doping level of $10^{18}$ cm$^{-3}$ and a thickness of 0.4 µm. For the n layer, we assumed a doping level of $10^{16}$ cm$^{-3}$ and a thickness of 2 µm. The photovoltaic sub-device layers' thicknesses were optimized for maximum efficiency by simulating the operation of the InAs thermophotovoltaic device with the silicon thermal emitter over the gap range considered in this study. We used the zinc and germanium dopant energy levels in the p and n layers, respectively. For the PV back contact, we considered a gold electrode, which also reflects the unabsorbed photons back to the PT emitter.

For both the thermionic emitter and the PV sub-device, the semiconductor materials' properties such as spectral absorptivity and electron and hole mobilities were taken from various empirical models considering their temperature and doping dependencies [53–56]. The various recombination coefficients and carrier lifetimes in Si and InAs were taken from the literature [57–60]. The dielectric permittivities of Si and InAs were taken from [53,61]. The dielectric permittivities of tungsten and gold were taken from [15,62]. The density of states and conductivity effective masses for the materials were taken from [60]. The temperature dependence of the effective density of states was considered. The temperature coefficients of the bandgap narrowing effect for Si and InAs were taken from [60]. The equilibrium Fermi levels in the PT emitter and the p and n regions of the PV sub-device were calculated using the charge neutrality criterion [45]. For ease of access, a list of the relevant material properties and device parameters is given in table 1.

**Table 1.** Material and device parameters used in the study

| Parameters | Si | InAs |
|---|---|---|
| Electron Auger recombination coefficient ($cm^6 s^{-1}$) | $1.1 \times 10^{-30}$ | $1.13 \times 10^{-27}$ |
| Hole Auger recombination coefficient ($cm^6 s^{-1}$) | $3 \times 10^{-31}$ | $1.13 \times 10^{-27}$ |
| Shockley-Reed-Hall lifetimes (s) | **[57] | **[59] |
| Front surface recombination velocity ($cm\ s^{-1}$) | 100 | 100 |
| Back surface recombination velocity ($cm\ s^{-1}$) | 0 | 0 |
| Temperature dependence of bandgap narrowing effect ($eVK^{-1}$) | $4.73 \times 10^{-4}$ | $2.76 \times 10^{-4}$ |
| Bandgap at 0 K (eV) | 1.17 | 0.415 |
| Electron density of states effective mass | 1.18 | 0.023 |
| Hole density of states effective mass | 0.81 | 0.41 |
| Carrier mobility | **[54] | **[55] |
| Dopant ionization energy (eV) | 0.045 | p type: 0.01<br>n type: 0.001 |
| Doping concentration ($cm^{-3}$) | $10^{18}$ | p type: $10^{18}$<br>n type: $10^{16}$ |
| Thickness (μm) | 20 | p type: 0.4<br>n type: 2 |
| Optical absorption coefficient | **[56] | **[53] |
| Richardson constant ($A\ cm^{-2} K^{-2}$) | 120 | |
| Vacuum gap size range (μm) | 0.1–10 | |
| Solar concentration ratio | 100 | |

Note: ** Dependencies were modelled from the cited references.

Here it is worth highlighting some of the dependencies of device performance on the parameters shown in Table 1. Note that, some of these trends are hypothetical as they assume the variation of a single parameter. However, in reality, changes in one parameter could also significantly affect other material parameters. (1) A higher solar concentration would lead to higher power density and device temperature. However, such an increase would not be proportional and at a very high solar concentration, the increment in power density and temperature (given that it doesn't exceed the melting point of the emitter material) might be very small due to the significant increase of the various loss processes. (2) A lower or higher interelectrode gap size (outside the range in table 1) would reduce the device performance due to the increase in radiative and/or space charge loss. (3) Higher recombination coefficients or surface recombination velocities would reduce the steady-state photogeneration of the electron-hole pairs and would decrease the photogeneration related performance [63]. (4) A lower bandgap for the photovoltaic device may not significantly improve the conversion performance and considering the thermal limitation and output voltage, it might even lead to lower overall performance.

## 4. Results and Discussion

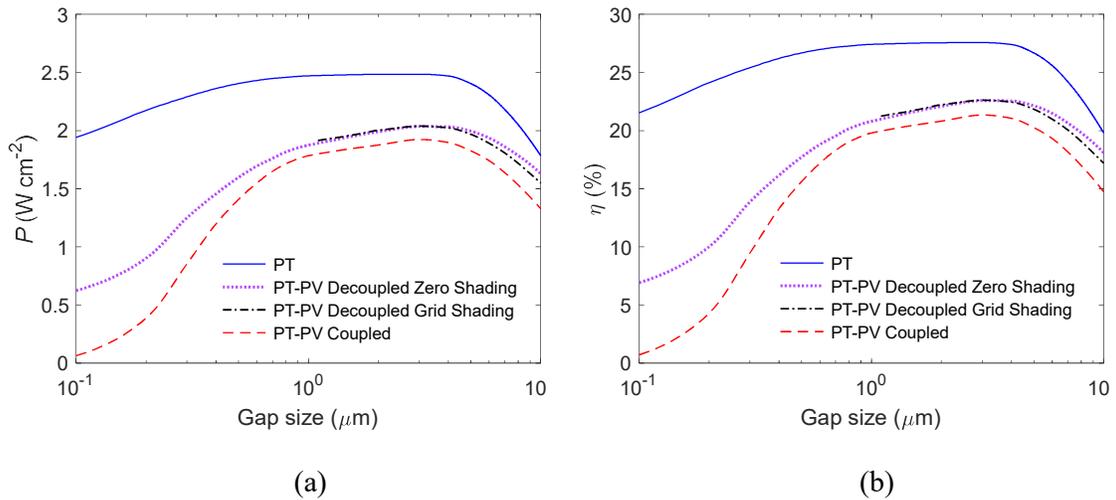

(a)                                             (b)

Fig. 3. (a) Output power density and (b) Conversion efficiency of various PT and PT-PV hybrid solar devices as a function of the vacuum gap size. The data are shown at MPP and for a solar concentration ratio of 100X.

Fig. 3 shows the output power density and conversion efficiency of various PT-PV hybrid solar devices. The decoupled devices provide higher performance compared to the coupled one. This is because the performance of the latter is limited by the constraint of current matching between its sub-devices. The additional electrode of the decoupled devices represents an extra degree of freedom that relieves this constraint and allows the PT and PV sub-devices to be biased at their respective maximum power points (MPPs). However, this comes at the additional costs of optical shading loss and voltage drop in the electrode. Nevertheless, this electrode, overall, could be advantageous for hybrid devices as evident from fig. 3. Among the decoupled devices, the one with a grid electrode provides slightly lower performance (compared to its grid-less counterpart) at larger gap sizes. This is due to the decrease in electronic and photonic coupling strengths caused by the grid structure. In order to place these results in a broader context, fig. 3 also shows the same metrics for a single PT device with a metallic collector. Interestingly, this device shows better performance compared to its hybrid counterparts. This might at first appear unexpected, but the reason may be traced to the high optical reflectivity of the metallic collector, leading to higher emitter temperature.

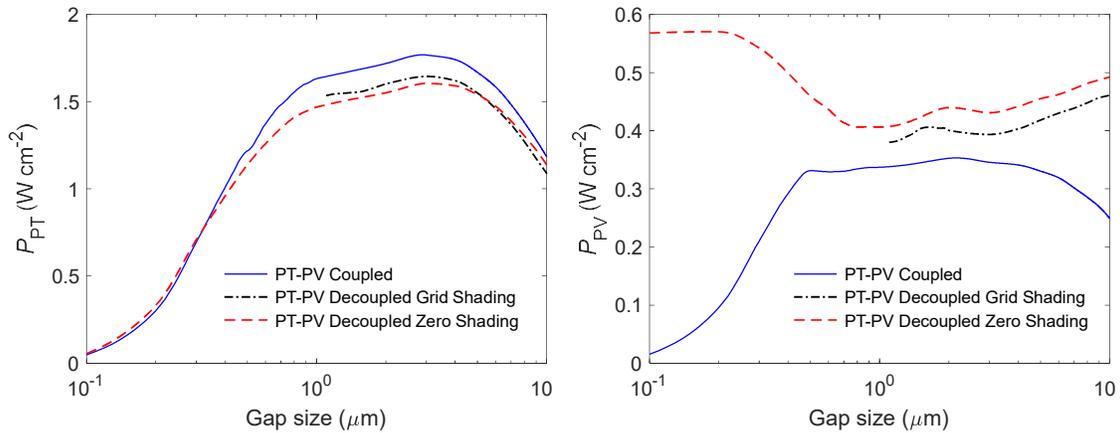

(a)                                                      (b)

Fig. 4. Output power density of the (a) PT and (b) PV sub-devices of various PT-PV hybrid solar devices as a function of the vacuum gap size. The data are shown at MPP and for a solar concentration ratio of 100X.

Fig. 4 shows the gap size trends of the PT and PV power densities for different hybrid solar devices. For the PT sub-devices, irrespectively of the device configuration, the power density trend shows a local maximum, which is due to the varying degrees of electronic and photonic coupling (between the PT emitter and the PV sub-device) as the gap size changes. In more detail, for small gaps, the photonic coupling is significantly enhanced due to photon tunnelling (also known as the near-field effect), which, in turn, prevents the emitter temperature from increasing. On the other hand, for such gaps, the space charge effect is negligible; however, the reduced temperature also precludes a strong electronic coupling. For larger gaps, the space charge effect is more severe and results in a weaker electronic coupling. On the other hand, despite the insignificance of the near-field effect for such gaps, the photonic coupling becomes stronger due to the increase in the PT emitter temperature.

The PT power density of the decoupled devices is lower than that of the coupled one over most of the gap range. This is counterintuitive given the additional degree of freedom offered by the additional electrode. However, this can be explained by the extent of space charge loss in various PT sub-devices. As will be seen later, under optimal conditions, the decoupled PT sub-devices operate at a higher emitter temperature. This exacerbates the space charge effect and shifts the optimal bias point to a higher voltage. This decreases the current density of the decoupled PT sub-devices (see fig. 5a) to such an extent that the power density also decreases.

The PV power density trend of the coupled device is qualitatively similar (due to the current matching constraint) to its PT counterpart. On the other hand, the PV power density trend of the grid-less decoupled device exhibits a local minimum. This is due to the strong dependence of the photonic coupling strength (between the PT emitter and the PV sub-device) on the gap size as explained before. The decoupled PV sub-devices, irrespectively of the top electrode structure, show a higher power density than the coupled one. However, the PV sub-device with a grid electrode shows lower performance (due to optical shading) than the grid-less one at identical gap sizes. The improved performance of the PV sub-device in the decoupled devices compensates for the performance loss in its PT counterpart and results in higher overall performance (compared to the coupled device) as shown in fig. 3.

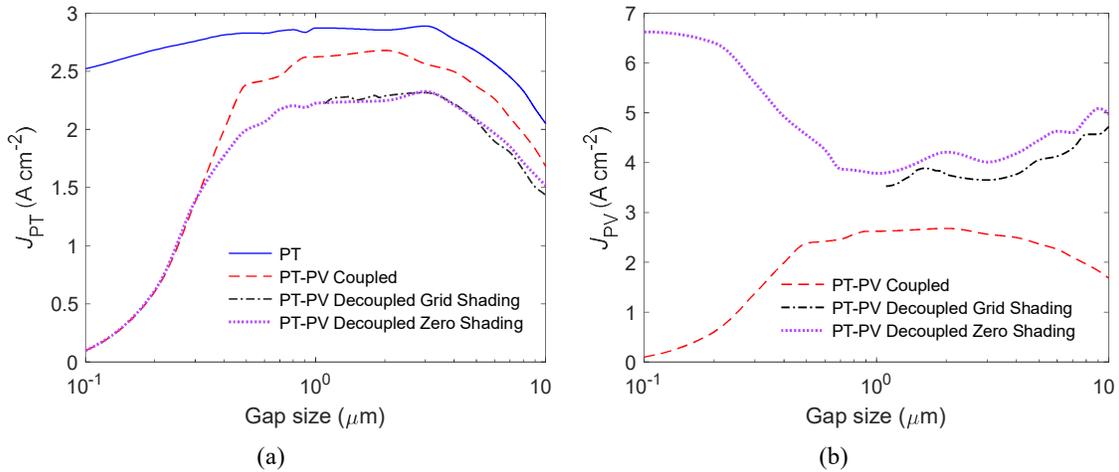

Fig. 5. Output current density of the (a) PT and (b) PV sub-devices of various PT and PT-PV hybrid solar devices as a function of the vacuum gap size. The data are shown at MPP and for a solar concentration ratio of 100X.

Fig. 5 shows the gap size trends of the thermionic and photovoltaic current densities. The general trends of the thermionic current density in various devices are qualitatively similar. This is again due to the variations of the electronic and photonic coupling strengths with gap size as explained above. Crucially, the PT device with a metallic collector provides a higher thermionic current compared to various hybrid solar devices, thanks to its higher emitter temperature due to the weaker photonic coupling with the collector in this device.

In the coupled device, we see that the PT sub-device is not constrained by the current matching at very small gaps. However, as the emitter temperature increases with the gap size (as will be seen later in fig. 6), current matching does eventually pose a constraint for the PT sub-device. Nevertheless, the coupled PT sub-device shows a higher current density compared to that of the decoupled devices at larger gaps. We attribute this to a more severe space charge loss in the decoupled devices as explained before. On the other hand, the thermionic current densities of the two decoupled devices are almost identical in their common gap size range.

For the photovoltaic current density, in the coupled device, the PV and PT sub-devices have to match their currents. However, in the grid-less decoupled device, the PV current density trend shows a local minimum. The initial decrease in current density is due to the reduced photonic coupling strength (*i.e.,* the weakening of the near-field effect) as the gap size increases. However, at larger gaps, this decrease in photonic coupling is countered by the increase in the PT emitter temperature; as a result, the current density increases again. The PV current density trend of the device with a grid electrode can be explained similarly, although grid shading lowers the current overall.

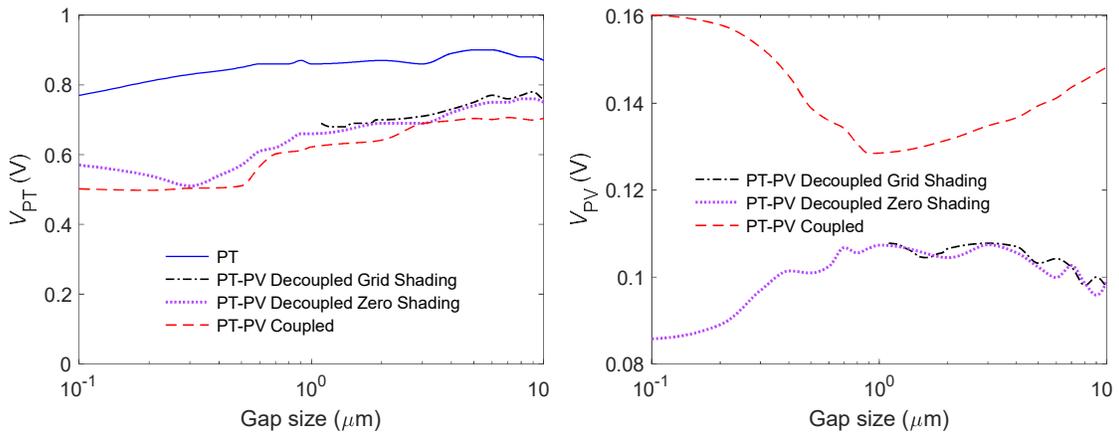

(a) (b)

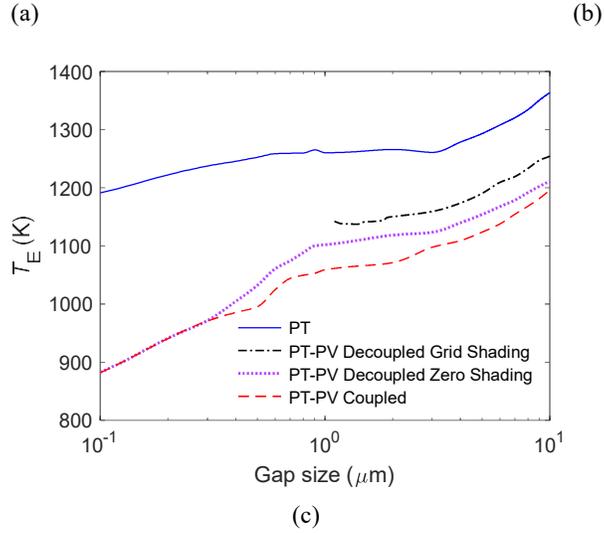

(c)

Fig. 6. Output voltage of the (a) PT and (b) PV sub-devices of various PT and PT-PV hybrid solar devices as a function of the vacuum gap size. (c) Emitter temperature of various PT and PT-PV hybrid solar devices. The data are shown at MPP and for a solar concentration ratio of 100X.

Fig. 6(a)-(b) show the voltage contributions from the thermionic and photovoltaic stages, respectively, for different hybrid solar devices. For reference, in fig. 6(a), we also show the output voltage of the PT solar device. Fig. 6(c) shows the respective PT emitter temperatures. We see a strong correlation between the PT sub-devices' voltages and emitter temperatures as the gap size increases. The coupled PV sub-device's voltage trend shows a local minimum (due to current matching). On the other hand, at a given gap size, the decoupled PV devices' output voltage is significantly reduced (due to a voltage drop in the series resistance associated with the additional electrode). Moreover, the voltage does not show a monotonic trend. This is due to the tradeoff between the PV current density and the resulting ohmic loss in the series resistance.

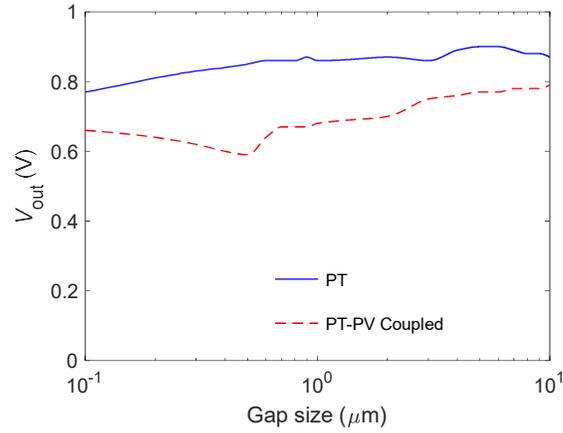

Fig. 7. Output voltage of the PT and PT-PV coupled hybrid solar devices as a function of the vacuum gap size. The data are shown at the MPP and for a solar concentration ratio of 100X.

To further explore the voltage gain from a direct electronic coupling between the PT and PV sub-devices, in fig. 7, we compare the total output voltage of the coupled PT-PV solar device with that of the single PT solar device. Crucially, despite the voltage boost from the PV device, the output voltage in the coupled device is significantly lower than that of the PT solar device. We attribute this to the stronger photonic coupling (resulting in a lower emitter temperature) between the coupled PT emitter and PV sub-device, and their current matching constraint.

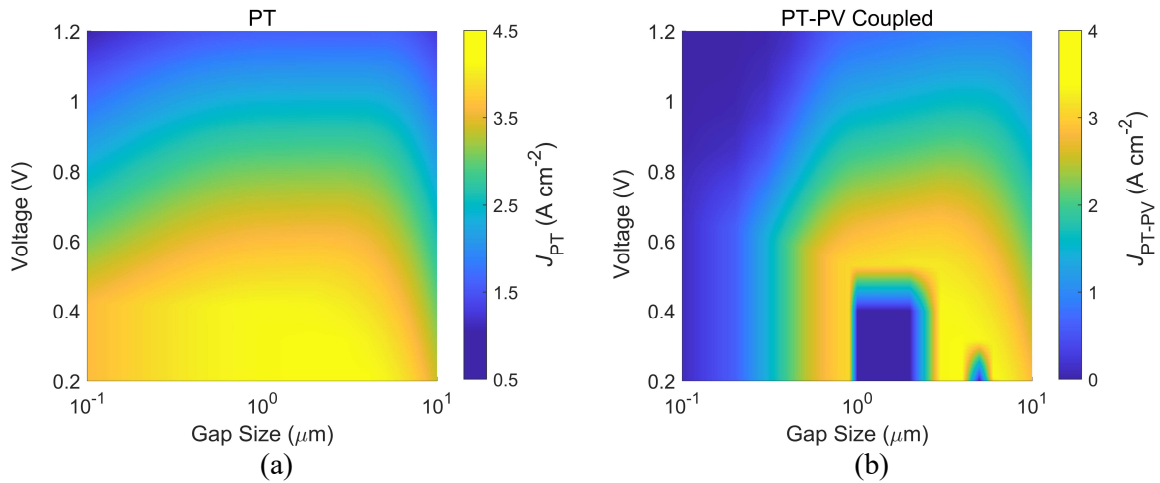

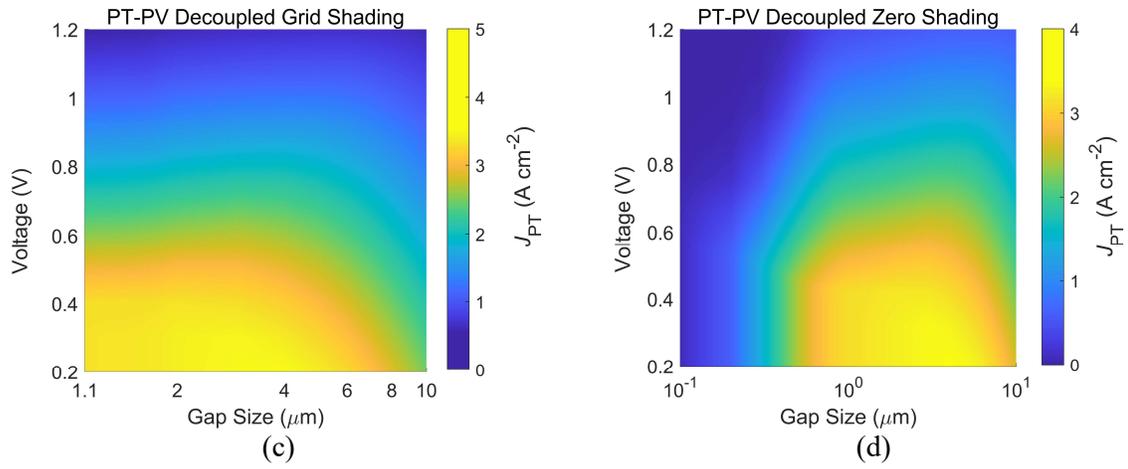

(c)                          (d)

Fig. 8. Photothermionic current density of (a) PT, (b) PT-PV coupled, PT-PV decoupled solar device (c) with grid electrode, and (d) without grid electrode, respectively as a function of the output voltage and gap size. The data are shown for a solar concentration ratio of 100X. Note that for the decoupled configurations, the voltage corresponds to that of the PT-sub device.

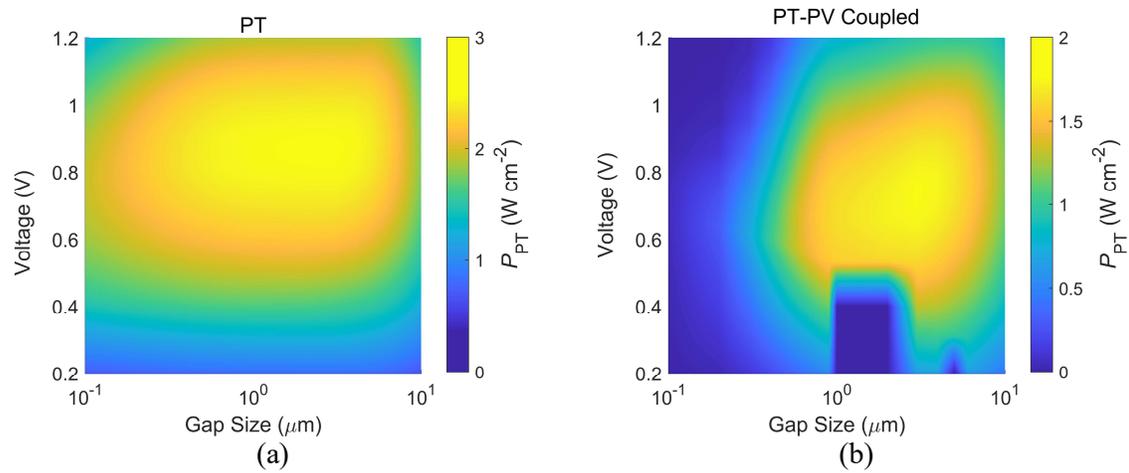

(a)                          (b)

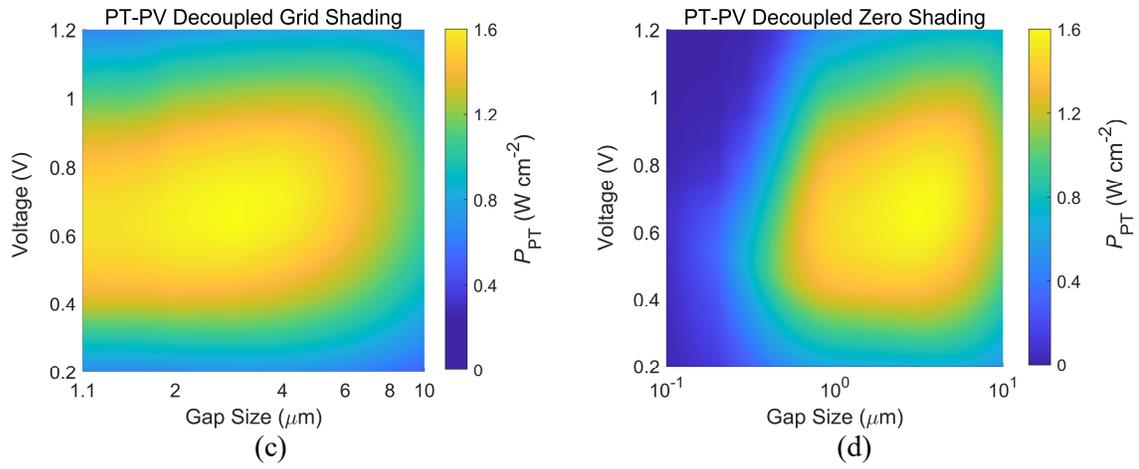

Fig. 9. Photothermionic power density of (a) PT, (b) PT-PV coupled, PT-PV decoupled solar device (c) with grid electrode, and (d) without grid electrode, respectively as a function of the output voltage and gap size. The data are shown for a solar concentration ratio of 100X. Note that for the decoupled configuration, the voltage corresponds to that of the PT-sub device.

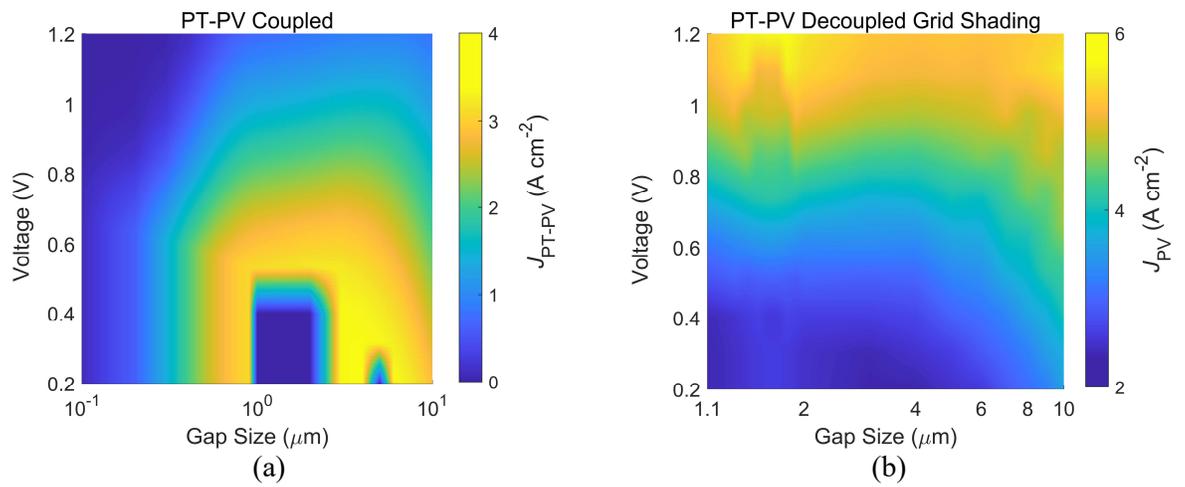

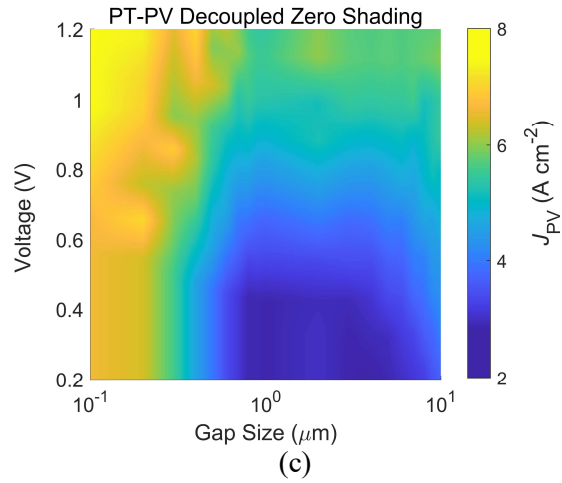

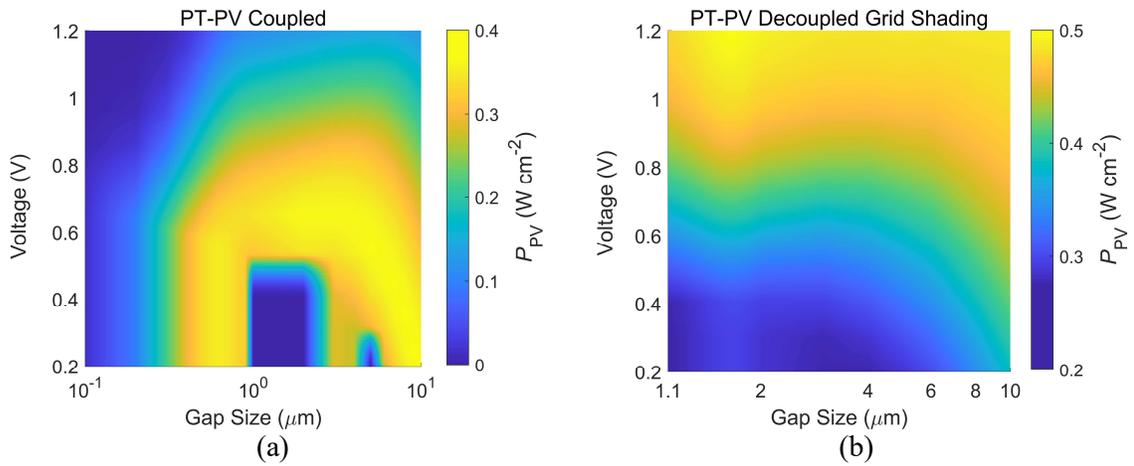

Fig. 10. Photovoltaic current density of (a) PT-PV coupled, PT-PV decoupled solar device (b) with grid electrode, and (c) without grid electrode, respectively as a function of the output voltage and gap size. The data are shown for a solar concentration ratio of 100X. Note that for the decoupled configuration, the voltage corresponds to that of the PT-sub device.

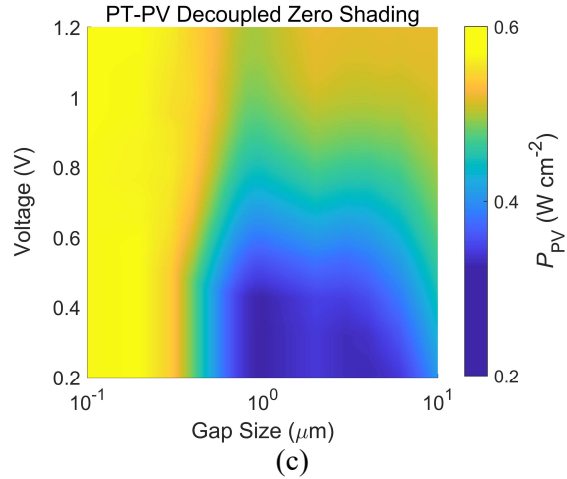

Fig. 11. Photovoltaic power density of (a) PT-PV coupled, PT-PV decoupled solar device (b) with grid electrode, and (c) without grid electrode, respectively as a function of the output voltage and gap size. The data are shown for a solar concentration ratio of 100X. Note that for the decoupled device, the voltage corresponds to that of the PT-sub devices.

To further illustrate the complex dependencies of the device operation on voltage and gap size, in figs. 8-11, we show the current and power density graphs of the PT and PV sub-devices as a function of these two variables. For a given voltage, the plotted quantities' trends with gap size can be explained by similar reasonings as discussed before. On the other hand, for a given gap size, the plotted quantities' variation with the voltage can be understood as follows. In general, for a given gap, as we increase the voltage, the electrons in the PT emitter need to overcome a higher vacuum barrier for thermionic emission, which leads to weaker electronic coupling between the emitter and collector (see fig. 8). As a result, the PT emitter temperature, which is dictated by thermal balance, increases with the output voltage[14]. This, however, also leads to stronger photonic coupling. Consequently, the PV power and current densities gradually increase with the voltage in the decoupled devices (see figs. 10-11). On the other hand, in the coupled device, the PV power and current densities (see figs. 10-11) decrease with the increase in voltage, which is due to the current matching constraint. Irrespectively to the device configuration, the voltage trend of the PT power density (see fig. 9) shows a local maximum, which is due to the tradeoff between the voltage and current density in the thermionic sub-devices. Crucially, for certain combinations of gap size and expected voltage in the coupled device, the current between the PT and PV sub-devices cannot be matched (see fig. 8), which leads to a zero-power density in both the thermionic and photovoltaic sub-devices (see figs. 9 and 11). Physically, this means that when a load is connected, the device will not be biased at that voltage.

5. Summary and Conclusion

In summary, this work presents a novel solar device by combining the photothermionic and photovoltaic mechanisms. The primary conceptual motivation behind this device is to utilize the solar spectrum more efficiently. The PT sub-device exploits the thermalization induced heating effect of the above-bandgap portion of solar energy; it may also experience a reduction in the effective work function through Fermi level splitting. However, due to high temperature operation, the PT sub-device also loses part of the absorbed solar energy via radiation. The photovoltaic stage can partly recycle this radiative loss in the PT sub-device as well as capture the solar photons which are not utilized by the former. We investigated the possible device architectures to implement such a hybrid solar device by considering various complex and interrelated physics. Our analysis shows that each architecture has its own set of challenges. For example, in the coupled configuration, the hybrid device performance is limited due to current matching between its sub-devices. In the decoupled configuration, the additional electrode relieves the current matching constraint, offering more flexibility in optimizing the individual thermionic and photovoltaic sub-devices, but introduces additional losses. We illustrated that the photonic and electronic coupling between the emitter and the collector strongly depend on the material's dielectric properties and play a decisive role in determining the device performance. Importantly, we found that a single photothermionic device with a metallic collector can outperform the hybrid device. In other words, the performance contribution from the PV stage in a hybrid device comes at the cost of lower emitter temperature due to a stronger photonic coupling between the emitter and the PV device. This temperature reduction ultimately creates a bottleneck in the overall hybrid device performance. Therefore, unless the PV sub-device ensures an electronic and photonic coupling strength similar to that of a metallic collector or lower temperatures are desired for practical reasons, the hybrid operation may not be justified for applications.

**Acknowledgements**

We acknowledge funding from the Natural Sciences and Engineering Research Council of Canada (Grants No. RGPIN-2017-04608 and No. RGPAS-2017-507958). This research was undertaken thanks in part to funding from the Canada First Research Excellence Fund, Quantum Materials and Future Technologies Program. Ehsanur Rahman thanks the Natural Sciences and Engineering Research Council of Canada for a Vanier Canada Graduate Scholarship and the University of British Columbia for an International Doctoral Fellowship and Faculty of Applied Science Graduate Award.

E.R. conceptualized the work; developed the algorithm; performed the simulation, model verification, and data visualization; and wrote the draft manuscript; A.N. provided technical guidance and reviewed and edited the manuscript.

to energy conversion. Int J Energy Res 2009;33:1203–32. https://doi.org/10.1002/er.1607.

[43] Francoeur M, Pinar Mengüç M, Vaillon R. Solution of near-field thermal radiation in one-dimensional layered media using dyadic Green's functions and the scattering matrix method. J Quant Spectrosc Radiat Transf 2009;110:2002–18. https://doi.org/10.1016/j.jqsrt.2009.05.010.

[44] Lough BC, Lewis RA, Zhang C. Principles of charge and heat transport in thermionic devices. In: Al-Sarawi SF, editor. Smart Struct. Devices, Syst. II, vol. 5649, SPIE; 2005, p. 332. https://doi.org/10.1117/12.582101.

[45] Sze SM. Semiconductor Devices: Physics and Technology. second. John Wiley & Sons, New York; 2002.

[46] Francoeur M, Vaillon R, Meng MP. Thermal impacts on the performance of nanoscale-gap thermophotovoltaic power generators. IEEE Trans Energy Convers 2011;26:686–98. https://doi.org/10.1109/TEC.2011.2118212.

[47] Vaillon R, Robin L, Muresan C, Ménézo C. Modeling of coupled spectral radiation, thermal and carrier transport in a silicon photovoltaic cell. Int J Heat Mass Transf 2006;49:4454–68. https://doi.org/10.1016/j.ijheatmasstransfer.2006.05.014.

[48] Koeck FAM, Nemanich RJ, Lazea A, Haenen K. Thermionic electron emission from low work-function phosphorus doped diamond films. Diam Relat Mater 2009;18:789–91. https://doi.org/10.1016/j.diamond.2009.01.024.

[49] Callahan WA, Feng D, Zhang ZM, Toberer ES, Ferguson AJ, Tervo EJ. Coupled Charge and Radiation Transport Processes in Thermophotovoltaic and Thermoradiative Cells. Phys Rev Appl 2021;15:1.